%% file: paper.tex
\documentclass{sig-alternate}

\input{preamble}

\usepackage[backend=bibtex,natbib=true,sortcites=true,style=numeric,maxcitenames=1,maxbibnames=100]{biblatex}

\renewcommand{\cite}{\citep}
\title{Pending}

\numberofauthors{2} 
\author{
\alignauthor
Eduardo Graells-Garrido\\
       \affaddr{Telefónica I+D}\\
       \affaddr{Santiago, Chile}\\
\alignauthor
Diego Saez-Trumper\\
       \affaddr{EURECAT}\\
       \affaddr{Barcelona, Spain}\\
}

\begin{document}

\title{A Day of Your Days: Estimating Individual Daily Journeys Using Mobile Data to Understand Urban Flow}

\maketitle

\begin{abstract}
\input{abstract.tex}
\end{abstract}

\input{paper_content.tex}

\printbibliography

\end{document}

%% file: preamble.tex
\usepackage{times}

\usepackage[utf8]{inputenc}
\usepackage{url}
\makeatletter
\def\url@leostyle{%
  \@ifundefined{selectfont}{\def\UrlFont{\sf}}{\def\UrlFont{\small\bf\ttfamily}}}
\makeatother
\usepackage{microtype}
\usepackage{graphicx}
\usepackage{booktabs}
\usepackage{tabulary}
\usepackage{subcaption}
\usepackage{amsmath}

\usepackage{enumitem}
\setlist{nolistsep}

\usepackage{algorithmicx}
\usepackage{algorithm}
\usepackage{algpseudocode}

\newfont{\mycrnotice}{ptmr8t at 7pt}
\newfont{\myconfname}{ptmri8t at 7pt}
\let\crnotice\mycrnotice%
\let\confname\myconfname%

\newcommand{\ie}{\emph{i.\,e.}}

\newcommand{\eg}{\emph{e.\,g.}}

\usepackage{gensymb}

\newcommand{\spara}[1]{\smallskip\noindent{\bf #1.}}

\usepackage{xpatch}

\xpatchbibmacro{name:andothers}{%
  \bibstring{andothers}%
}{%
  \bibstring[\emph]{andothers}%
}{}{}

\usepackage[draft]{hyperref}

\toappear{Under review--do not distribute. Contact the authors before citing.}

%% file: abstract.tex
Nowadays, travel surveys provide rich information about urban mobility
and commuting patterns. But, at the same time, they have drawbacks: they
are static pictures of a dynamic phenomena, are expensive to make, and
take prolonged periods of time to finish. However, the availability of
mobile usage data (\emph{Call Detail Records}) makes the study of urban
mobility possible at levels not known before. This has been done in the
past with good results--mobile data makes possible to find and
understand aggregated mobility patterns. In this paper, we propose to
analyze mobile data at individual level by estimating \emph{daily
journeys}, and use those journeys to build Origin-Destiny matrices to
understand urban flow. We evaluate this approach with large anonymized
CDRs from Santiago, Chile, and find that our method has a high
correlation ($\rho = 0.89$) with the current travel survey, and that it
captures external anomalies in daily travel patterns, making our method
suitable for inclusion into urban computing applications.

%% file: paper_content.tex
\section{Introduction}\label{introduction}

Travel surveys provide information about urban mobility and commuting
patterns, mainly through Origin-Destiny matrices derived from them.
These matrices allow urban planners and policy makers to understand
travel patterns in urban mobility. Such surveys are usually collected
once per decade, due to their expensiveness (both in time and monetary
costs). Moreover, they are limited in some ways: travel surveys are
static pictures of a dynamic phenomena and, due to its sample size, they
are limited to big areas (either administrative or designed).

In this paper, we propose to use mobile data used for billing, which
indicates a subset of the antennas a mobile device has connected through
the day, as well as the corresponding timestamps. We use these digital
footprints to build extended travel diaries. Travel diaries are the
basic elements of travel surveys, but we extend them through daily
journeys, as we detect not only trips, but also other ``non-trip''
activities. From these daily journeys we build Origin-Destiny (OD)
matrices, at a fraction of the expenses needed to build OD matrices from
travel surveys.

Our main contribution is a method to detect these disaggregated daily
journeys using \emph{Call Detail Records} (CDRs). Our approach is based
on \emph{graphical timelines} and computational geometry algorithms,
which are applied having in mind transport-based rules regarding trip
duration and distance. We use an anonymized CDR dataset from one of the
largest telecommunications company in Chile, with a market share of
38.18\% as of June 2015. Chile, being one of the developing countries
with highest mobile phone penetration, is a good candidate for analyzing
urban mobility using CDRs--for instance, there are 132 mobile
subscriptions per 100
inhabitants.\footnote{\url{http://www.subtel.gob.cl/estudios-y-estadisticas/telefonia/}}
Particularly, we focus on Santiago, its capital and most populated city.

To evaluate our results, we compare our predicted urban flow (in the
form of an OD matrix) with the last travel survey for Santiago,
performed during 2012--2013. In terms of OD pairs, we obtain a very high
correlation ($\rho = 0.89$), indicating that our method recognizes the
urban flow on the city. We apply our methods at different days, and find
that, in addition, we detect how urban flows change in the presence of
unexpected conditions. This, jointly with the disaggregated nature of
our method, has potential for applications in urban computing
\cite{zheng2014urban}, discussed at the end of this paper.

\section{Related Work}\label{related-work}

The estimation of OD matrices, and thus, urban mobility patterns, is not
new. The most basic way of estimating those patterns is by travel
surveys, but also other methods have been developed. A common method is
based on traffic counts \cite{cascetta1984estimation}, but the massive
availability of other kinds of information has allowed to estimate such
matrices in other ways, for instance, by using smart-card passive data
\cite{munizaga2012estimation} and, as in this paper, mobile data
\cite{gonzalez2008understanding,song2010limits,alexander2015origin,iqbal2014development,frias2012estimation}.

\begin{figure}[t]
\centering
\includegraphics[width=\linewidth]{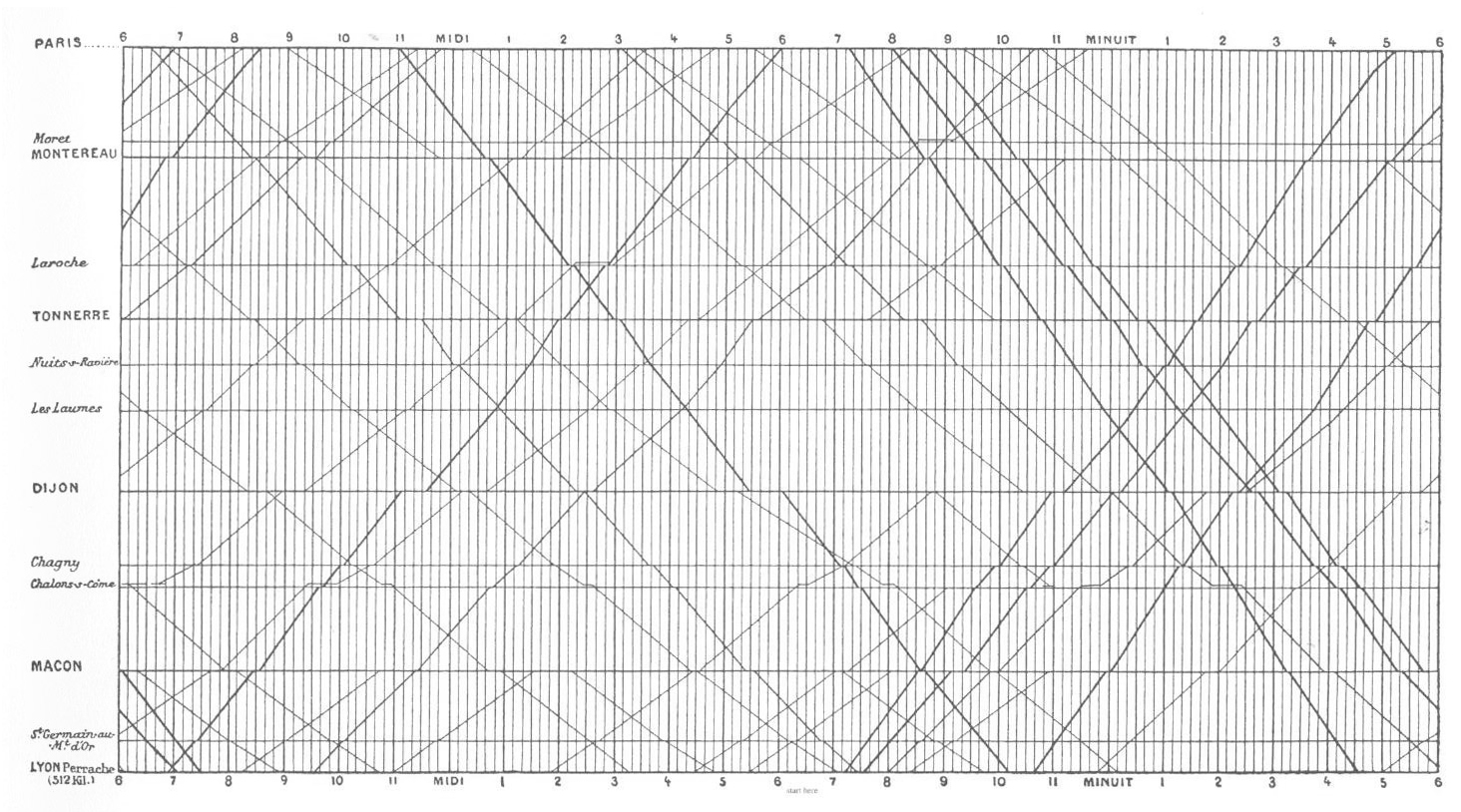}
\caption{Graphical timetable of a train schedule, by E.J. Marey, 1885.}
\label{fig:ejmarey}
\end{figure}

The work on mobile datasets has included both theory and practice. From
a theoretical point of view, the analysis of how predictable humans are
\cite{gonzalez2008understanding,song2010limits}. A practical work has
been the prediction of transient OD matrices by analyzing transitions of
connections between cell antennas. These transitions are the basis of
many methods, which employ techniques from optimization and temporal
association rules \cite{frias2012estimation}, to transport-based rules
to accept or discard transitions
\cite{iqbal2014development,alexander2015origin}. Our approach is also
based on antenna transitions. However, the main difference with previous
work is that we reconstruct individual \emph{daily journeys}, an
extended version of the travel diaries used to build travel surveys.
These journeys are built using a geometric approach based on
\emph{graphical timelines} \cite{tufte1993envisioning}, which are
time-space diagrams displaying timelines according to time and distance
covered. On these timelines we estimate the ``turning points'' of the
daily journey, which serve to differentiate daily activities into trips
and non-trips. These timetables are regularly used in transport to
display and analyze schedules, as well as phenomena associated with
vehicle behavior. For instance, Figure \ref{fig:ejmarey} shows a train
schedule designed in 1885 by Étienne-Jules Marey. On it, passengers can
see arrival and departure times, as well as the duration of stops and
the velocity of trains. Transport planners can study vehicle
behavior--since all vehicles share the same origin, if we were
visualizing their true trajectories instead of the scheduled ones,
\emph{vehicle bunching} can be recognized immediately
\cite{moreira2012bus}.

\section{Methods}\label{methods}

Our methods consider \emph{mobile user traces} estimated from anonymized
CDR data. CDRs are comprised of logged data extracted from cell-phone
antennas, and are used to bill customers. They contain the following
events: \emph{calls}, \emph{SMS} (text messages), and \emph{data events}
(triggered every 15 Megabytes or every 15 minutes of an active
connection). The following features are common between all events, and
thus are considered for analysis: the anonymized user ID, the antenna
ID, and time of the day. The antenna ID is used to determine a likely
position for the user at the corresponding time of the day. Note that
this position is assumed to be the same for all phones connected to the
same antenna, \ie, we do not perform triangulation based on signal
strength with nearby antennas.

\spara{Problem Definition and Proposal} The problem we propose to solve
is the estimation of a diary of activities for a day, using CDRs from
mobile data. A daily journey $J$ for user $u$ is defined as follows:
\[J_u = \{(A_i, (t_{iO}, t_{iD}), (p_{iO}, p_{iD})\}.\] Where $A_i$ is
an activity, $p_i$ (and $t_i$) are the positions (and times) associated
to the start/origin and end/destination of $A_i$. Activities can be of
types \emph{trip}, \emph{non-trip}, and \emph{unknown}.

To solve this problem we propose a two-step algorithm: first, we define
the candidate turning points of a day, where turning point is a moment
in the day, at a specific position, where the user started to perform an
activity (and, by definition, ends performing a previous activity). The
second step is to detect activities based on the candidate turning
points.

\spara{Graphical Timetables and Turning Points} For all users, we build
the following vector:
\[\vec{u} = [(t_0, p_0), (t_1, p_1), \ldots, (t_n, p_n)],\] Where each
element in $\vec{u}$ corresponds to an event in the CDR events of $u$ in
a day, with the corresponding timestamp $t$ and the antenna position
$p$. These vectors can be transformed into graphical timetables (see
Figure \ref{fig:ejmarey}), where the x-axis is the elapsed time during
the day, and the y-axis is the traveled (accumulated) distance from the
starting point. The different groups of segments present in the
timetable define the activities in $J_u$.

\begin{figure}[t]
\centering
\includegraphics[width=0.38\linewidth]{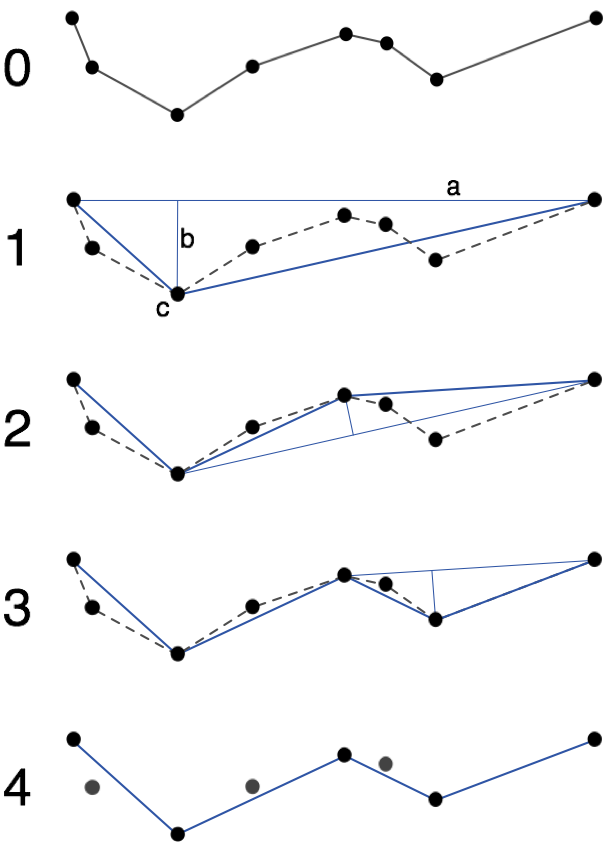}
\caption{Illustration of the Ramer-Douglas-Peucker algorithm. Source: Wikipedia Commons.}
\label{fig:rdp}
\end{figure}

However, due to the potential amount of CDR events, several contiguous
segments could be related to the same activity. To tackle this, we
propose to simplify the timetable using the Ramer-Douglas-Pecker line
simplification algorithm \cite{douglas1973algorithms}. This algorithm,
depicted on Figure \ref{fig:rdp}, starts with the extremes of the
trajectory, and iteratively adds vertices according to their distance to
the candidate simplified line. The vertex with greater distance (or
error) is added, and the process is repeated until the error is less
than a given tolerance. This results in a simplified vector $\vec{u_s}$,
which contains candidate ``turning points'' of daily activity. Those
candidate points define the potential activity segments that will end
defining $J_u$.

\spara{Activity Classification} The second step is to identify which
contiguous segments in the line defined by the points of $\vec{u_s}$ can
be merged into a single activity $A_i$. The most direct approach is to
merge a series of horizontal (or nearly horizontal) segments. However, a
continuously moving user (and thus, with non-horizontal segments) is
also feasible. Take the example of a taxi driver, whose primary activity
during the day is working as a driver. Although these activities are
valid, they are not relevant for an OD matrix, as they are not trips in
the individual sense. Thus, we define the following classifications:
\emph{unknown}, \emph{non-trips}, and \emph{trips}.

\emph{Unknown} are segments where the total distance covered is greater
than 100 kilometers. In those cases we cannot distinguish between trips
and unknown situations. For instance, the mobile number is associated to
a vehicle (\eg, a taxi) or another vehicle/device. While these are
indeed displacements in time and space, they do not fall on the daily
journey concept we study in this paper. Another case is when users
switch to WiFi networks, and thus disappear from the event log for a
long period of time. For instance, users who switch to the wireless
network of their work place might not be detected there, and the next
logged event could appear after working hours. On those cases, we cannot
identify trips reliably. To avoid this scenario, in some cases the
displacement between events has been limited to specific time windows
(\eg, 10 minutes and 1 hour \cite{iqbal2014development}).

\emph{Non-trips} are segments that are not \emph{unknown}. The criteria
to assign this type to an activity is based on the covered distance and
time. On the one hand, the antenna density of origin/destiny locations
of the activity is considered to determine a minimum distance that
cannot be attributed to signalling changes (this is discussed further in
the next section). On the other hand, some activities involve
displacements (\eg, working/studying in a big campus), but the speed of
movement is much slower than when performing a trip. Thus, if there is a
distance displacement, but the time is greater than 180 minutes, we
still consider a non-trip activity.

Finally, \emph{trips} are segments that are not \emph{unknown} and are
not \emph{non-trip}, but, in addition to the 180 minute rule, we enforce
a minimum duration of 15 minutes for a segment, due to the granularity
of the CDR data. Any other activity that does not fall into the rules
defined above is considered is considered \emph{unknown}.

\spara{Activity Merging} Having assigned an activity to each segment
built from $\vec{u_s}$, we procede to merge contiguous segments that
have the same activity. Two or more segments are merged into an activity
$A_i$ by considering the first time and position in the segment as
origin, and the last time and position in the segment as destination. We
also consider an additional case: when two \emph{trip} activities
surround a \emph{non-trip} activity, and the duration of the latter is
lesser or equal than 15 minutes, its activity is changed to \emph{trip}.
This scenario correspond to situations when users in public transport
make a connection, or users in vehicles face congested traffic. In this
way, after merging all activities, the daily journey $J_u$ is built.

\section{Context and Dataset}\label{context-and-dataset}

\begin{figure}[t]
\centering
\scriptsize
\includegraphics[width=0.99\linewidth]{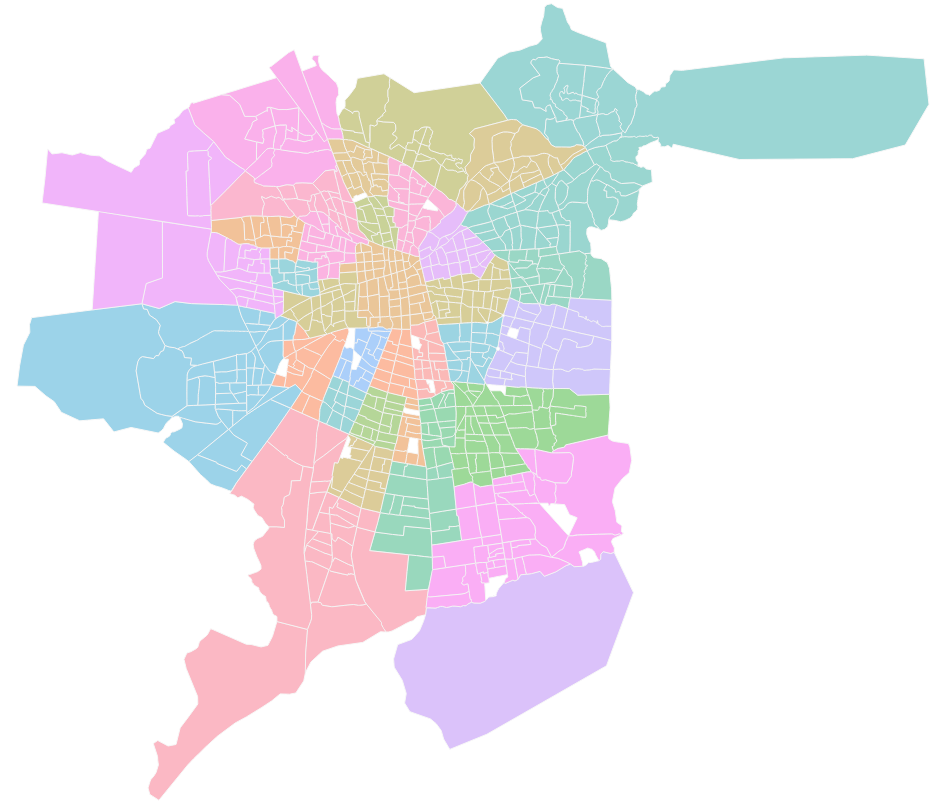}
\caption{The 752 zones from the OD Survey zoning. Colors indicate the 35 municipalities under consideration.}
\label{fig:stgo_municipalities}
\end{figure}

We work with a dataset from Santiago, the capital of Chile. Santiago is
a city with almost 8 million inhabitants, and it has an integrated
public transport system named Transantiago. The Metropolitan Area of
Santiago is composed of 35 independent administrative units named
municipalities. We work with this set of municipalities, depicted on
Figure \ref{fig:stgo_municipalities}.

\spara{Santiago 2012 Travel (OD) Survey} The Santiago 2012 travel survey
(\emph{ODS} hereafter) was performed during
2012--2013.\footnote{\url{http://www.sectra.gob.cl/biblioteca/detalle1.asp?mfn=3253}.}
It took almost one year to finish, and contains 96,013 trips (from
40,889 users). The information of trips is obtained through the travel
diaries fulfilled by the surveyed persons. The ODS, used to define
public policy related to public and private transport in the city, as
well as general urban mobility, is performed every 10 years due to its
costs and its difficulty.

The survey considers other municipalities outside the area, as well as
cities in other regions, due to the characteristics of the survey
procedure. Additionally, the survey also defines a zoning of the city,
with 752 zones within the considered municipalities. Each zone intends
to control for land-use and population density. Even though each trip in
the survey is associated to a zone and a municipality, the survey is
only representative at municipal level. At its current granularity, the
data available is not enough to calculate reliable mobility patterns at
zone level.

\begin{figure}[t]
\centering
\includegraphics[width=\linewidth]{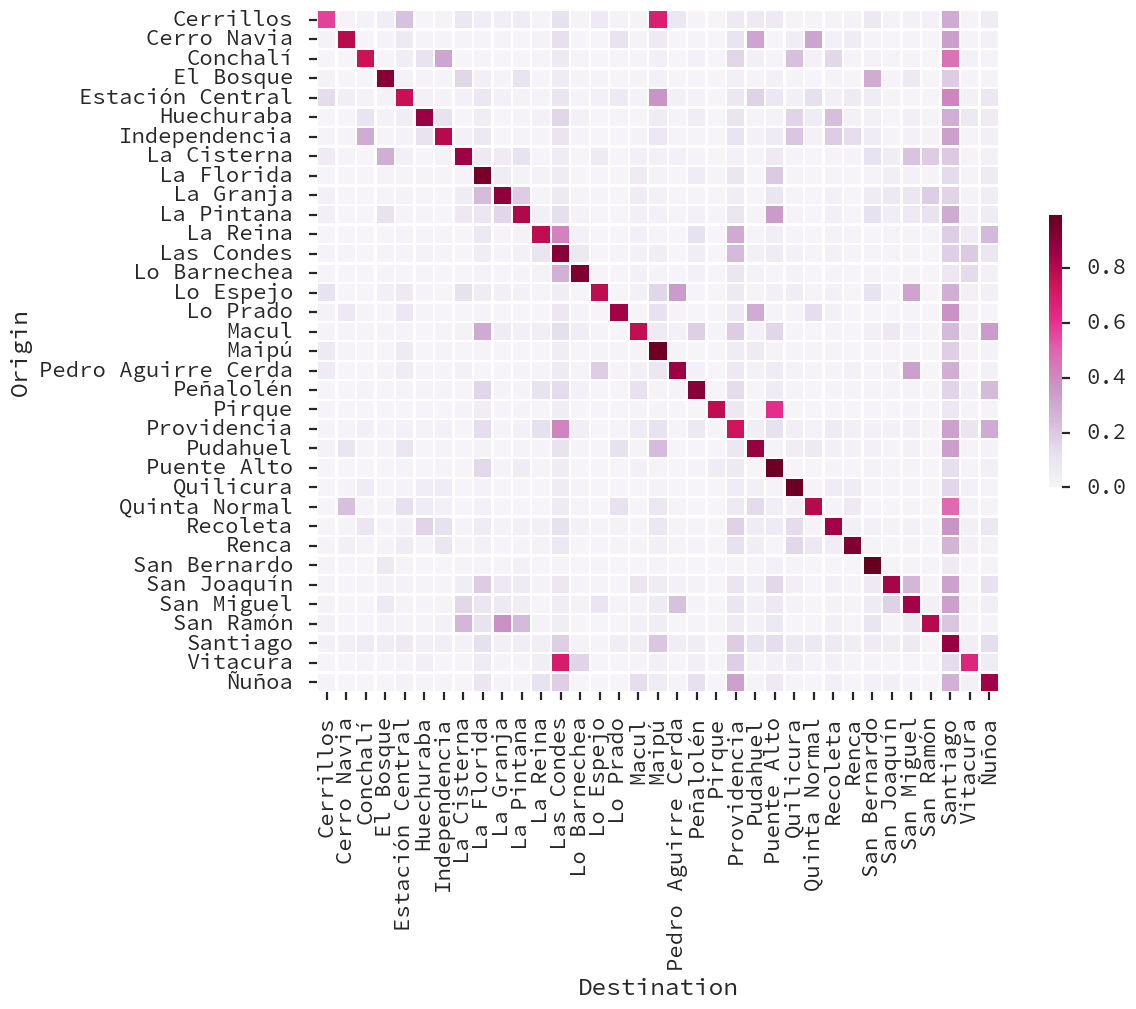}
\caption{Distributions of OD Survey 2012 trip data from Santiago, Chile. The matrix has been L2 normalized on rows (origins).}
\label{fig:eod_2012_matrix}
\end{figure}

In this paper, without losing generality, we focus on the 51,819 trips
performed on working days, from 22,541 users, performed on the inner 35
municipalities of the Metropolitan Area. We aggregated trips according
to municipality into an OD matrix, shown in Figure
\ref{fig:eod_2012_matrix}. One can see that the most common trips are
intra-municipalities, but there are still municipalities that tend to
receive more trips than others. This is because most of the commercial
and working land-use is on the municipalities of \emph{Santiago},
\emph{Providencia}, \emph{Las Condes} and \emph{Vitacura}. Note that the
municipality of Santiago is at the center of the Santiago Metropolitan
Area. In the rest of this paper, when we mention Santiago we refer to
the Metropolitan Area.

\begin{figure}[t]
\centering
\includegraphics[width=\linewidth]{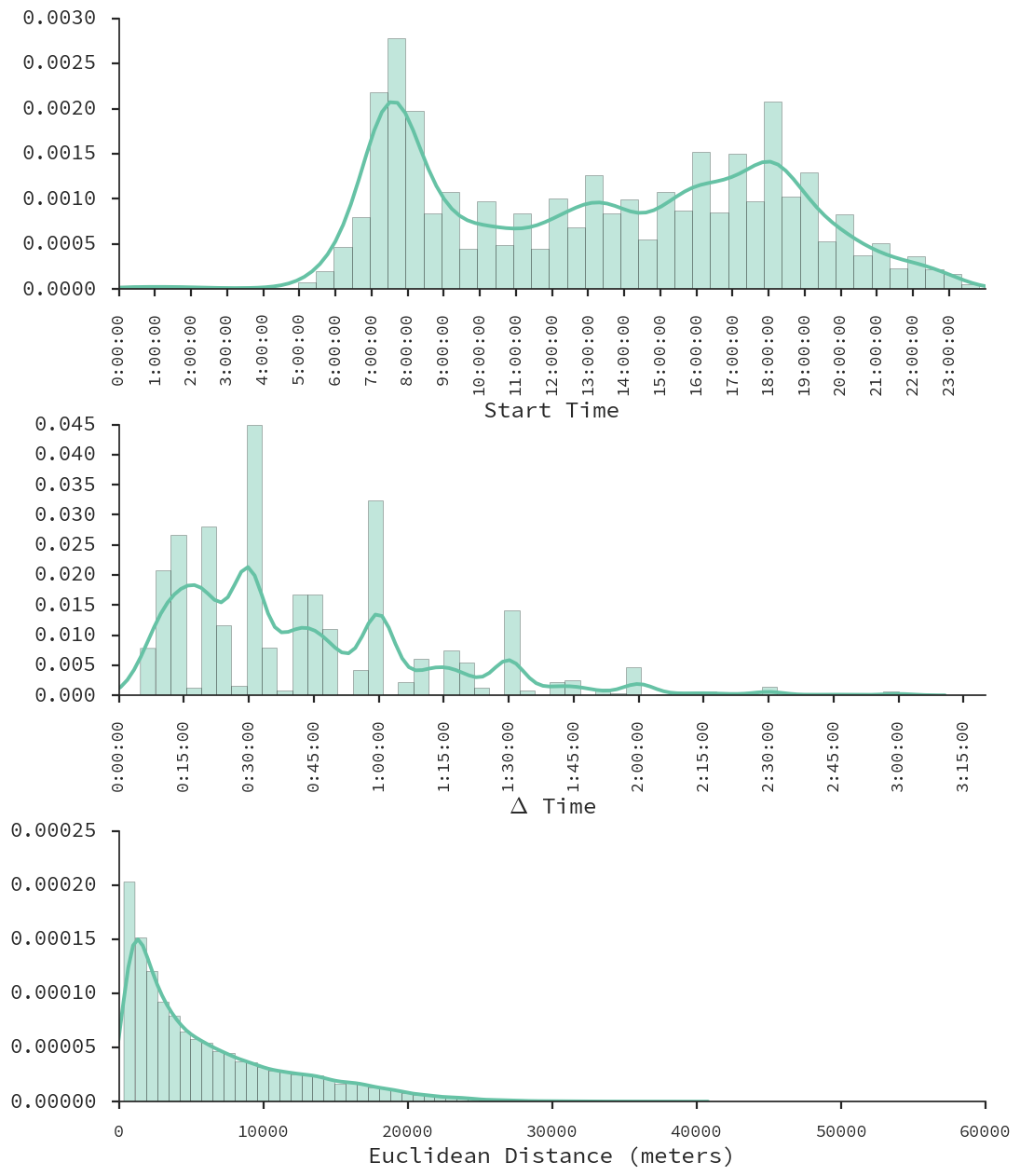}
\caption{Distributions of EOD Survey 2012 trip data.}
\label{fig:eod_2012_distributions}
\end{figure}

In terms of trip variables, Figure \ref{fig:eod_2012_distributions}
shows the distributions of trip start time, trip duration, and
approximated trip distance (\ie, euclidean distance). One can see that
trip start time follows an expected pattern of two high peaks (one in
the morning and one in the afternoon), with a third smaller peak at
lunch time. With respect to trip duration, the mean duration is 41
minutes. Note that the self-reported nature of the survey is evident,
due to the several peaks present on the distribution in 15 minute
periods (\eg, 30 and 45 minutes). Finally, with respect to travel
distance, the mean euclidean distance is 6.05 kilometers.

\begin{figure}[t]
\centering
\scriptsize
\includegraphics[width=0.99\linewidth]{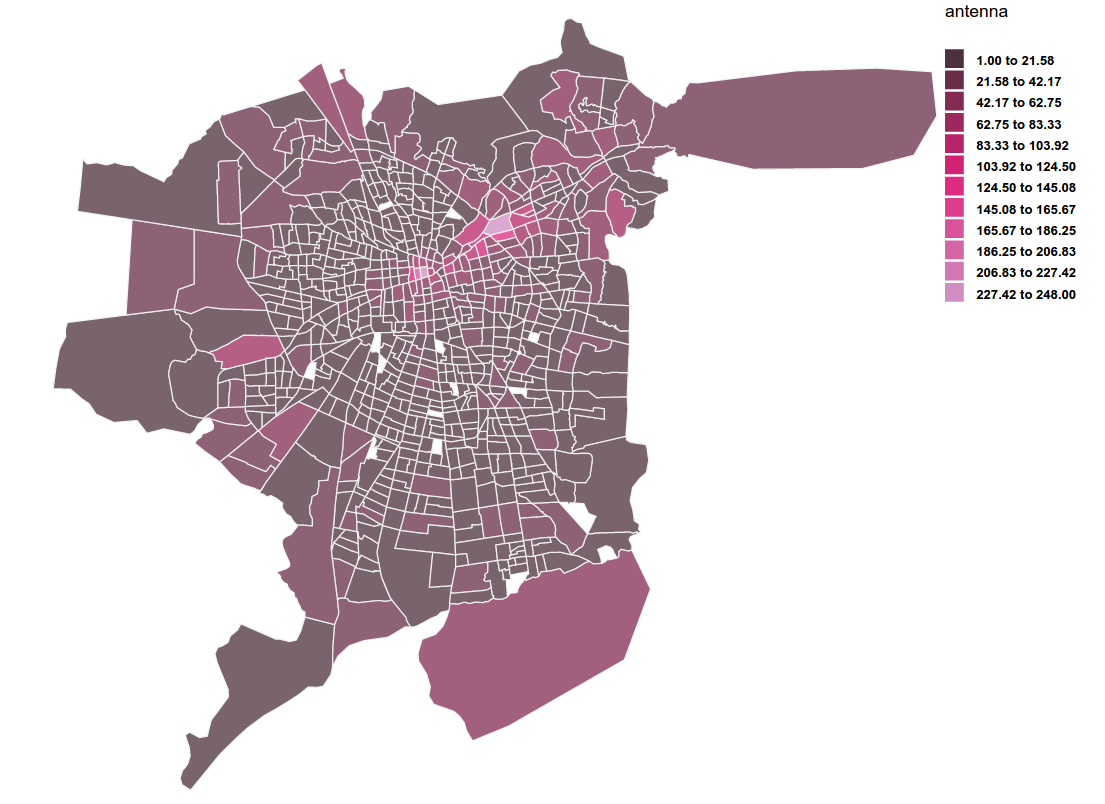}
\caption{Zoning of Santiago from the OD Survey. Colors indicate antenna density in each zone.}
\label{fig:stgo_antenna_coverage}
\end{figure}

\spara{Mobility Data} We analyze mobile data using CDRs from one of the
largest telecommunications company in Chile. The CDR data contains
events for all Mondays and Tuesdays of June 2015.\\In the 35
municipalities under consideration, the company has 12,936 antennas,
with 98\% of the zones having at least one antenna. Figure
\ref{fig:stgo_antenna_coverage} displays the antenna territorial
density. Note that the antenna distribution is not homogeneous on the
city, nor at any level. For instance, antenna distribution is correlated
with the ODS, considering aggregated destinations at municipal level
($\rho = 0.91$, $p < 0.001$).

To avoid artifacts in the trip detection introduced by connectivity
changes due to signal strength (\ie, mobile phones looking for the best
signal) and signal balancing (\ie, mobile antennas pointing devices to
connect to other antennas due to saturation), we estimated a distance
matrix $D_{z_i}$ for all antennas in each zone. Then, we defined that
the minimum distance for an activity to be considered a trip between
zones $z_o$ and $z_d$ is:
\[d_{\min} = \max \{ Q(D_{z_o}), Q(D_{z_d}) \}.\] Where $Q$ is the
quantile function (by manual experimentation we have found that $0.8$ is
a good value). The mean value of $Q(D_z)$ is 732 meters (minimum of 45
m., and maximum of 14.1 km.).

\begin{figure}[t]
\centering
\includegraphics[width=\linewidth]{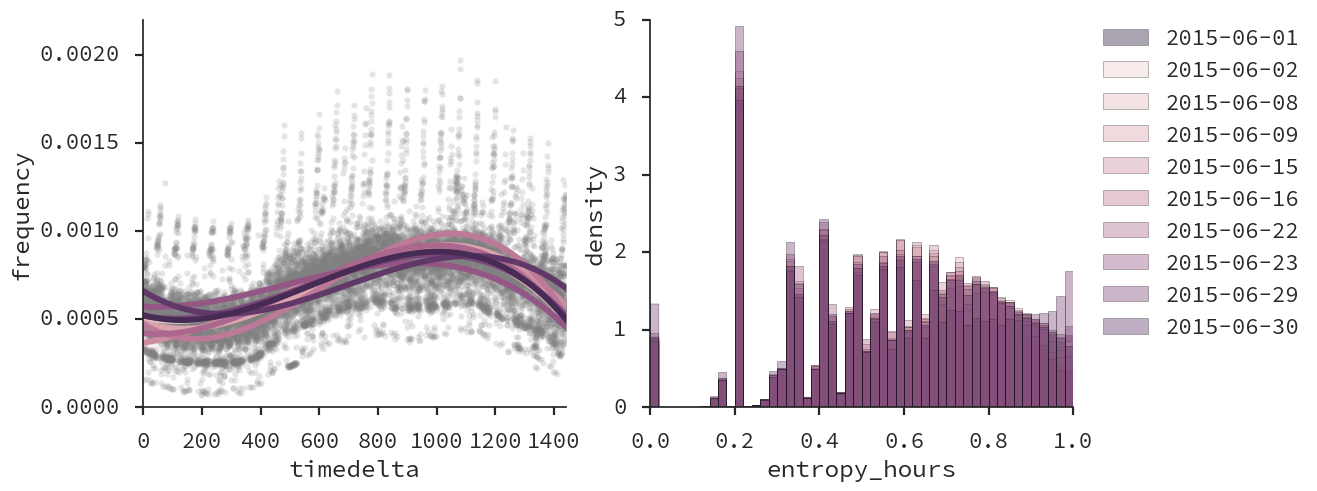}
\caption{Distributions of CDR event frequency (left) and entropy with respect to hours of the day (right).}
\label{fig:cdr_entropy_hours}
\end{figure}

While we do not disclose the exact number of users in each day due to
confidentiality and commercial issues, Figure
\ref{fig:cdr_entropy_hours} displays the distributions of event
frequency and hourly entropy for all days in the dataset. In the left
chart, each dot is a minute in a specific day. The frequency encodes the
fraction of events that the dot contains per day. One can see that the
distribution of frequency of events can be approximated by a cubic
linear regression, with a higher frequency of events in the afternoon.

The right chart of Figure \ref{fig:cdr_entropy_hours} displays the
distribution of user entropy with respect to hours of the day, for each
day. This entropy is defined as the Shannon entropy of user $u$:
\[H_u = - \sum p_{i,u} \ln p_{i,u}.\] Where $p_{i,u}$ is the probability
that user $u$ has a CDR event in the $i$th hour of the day. The purpose
of estimating this entropy is to have a measure of diversity with
respect to time for each user. Thus, we discard users who do not have
enough diversity to be able to estimate their journeys, or that have too
much diversity to be normal users (\eg, they could be SIM cards
associated to machines). We discarded users in the first quartile
($H_u < 0.4$) and in the last decile ($H_u > 0.9$).

We apply our methods to this dataset on the following section. To be
able to compare with the ODS, we employ a similar sample size of
$N = 100,000$ randomly selected mobile users (before filtering by
entropy).

\section{Daily Journeys and OD
Matrix}\label{daily-journeys-and-od-matrix}

\begin{figure}[t]
\centering
\includegraphics[width=\linewidth]{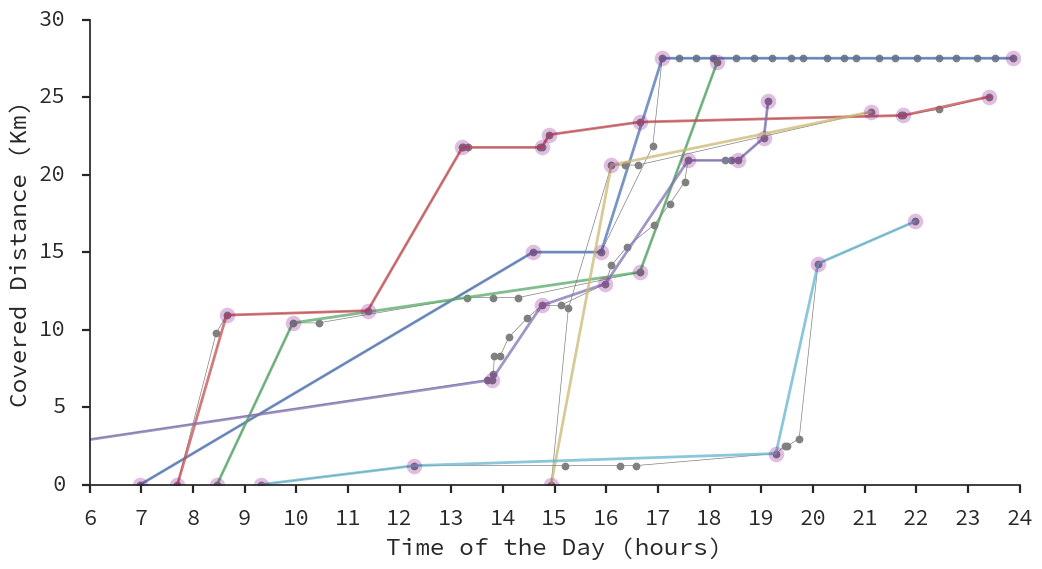}
\caption{Sample graphical timelines of mobile user traces with our method applied, including the RDP line simplification algorithm.}
\label{fig:algorithm_example}
\end{figure}

We applied our method to generate daily journeys $J_u$ to the CDR
dataset. Figure \ref{fig:algorithm_example} shows the result of randomly
selected samples from the dataset. Each colored line is a different
user, and the grey line underneath each colored line is the original
timetable before simplification. The remaining ``turning points'' are
rendered in purple, with a bigger size for easy identification on the
image. Those points define the activity segments which we classify as
\emph{unknown}, \emph{non-trip} and \emph{trip} according to the
definitions provided earlier.

\begin{table}[t]
\scriptsize
\centering
\begin{tabulary}{\linewidth}{lRRRR}
\toprule
Date &      N &  \# Trips &  Mean Time (m) &  Mean Distance (km) \\
\midrule
2015-06-01 &  37,037 &    56,148 &          78.38 &                7.68 \\
2015-06-02 &  37,622 &    57,764 &          77.78 &                7.74 \\
2015-06-08 &  34,272 &    50,017 &          83.86 &                7.80 \\
2015-06-09 &  34,611 &    50,778 &          83.03 &                7.77 \\
2015-06-15 &  36,274 &    53,642 &          79.45 &                7.68 \\
2015-06-16 &  36,804 &    56,046 &          78.16 &                7.68 \\
2015-06-22 &  34,288 &    49,434 &          80.05 &                7.78 \\
2015-06-23 &  36,239 &    54,983 &          77.62 &                7.71 \\
2015-06-29 &  22,770 &    31,963 &          74.78 &                7.50 \\
2015-06-30 &  35,181 &    52,597 &          78.94 &                7.66 \\
\bottomrule
\end{tabulary}
\caption{Number of users and trips for each day analyzed, as well as mean trip duration and mean euclidean distance.}
\label{table:cdr_means}
\end{table}

\begin{figure}[t]
\centering
\includegraphics[width=\linewidth]{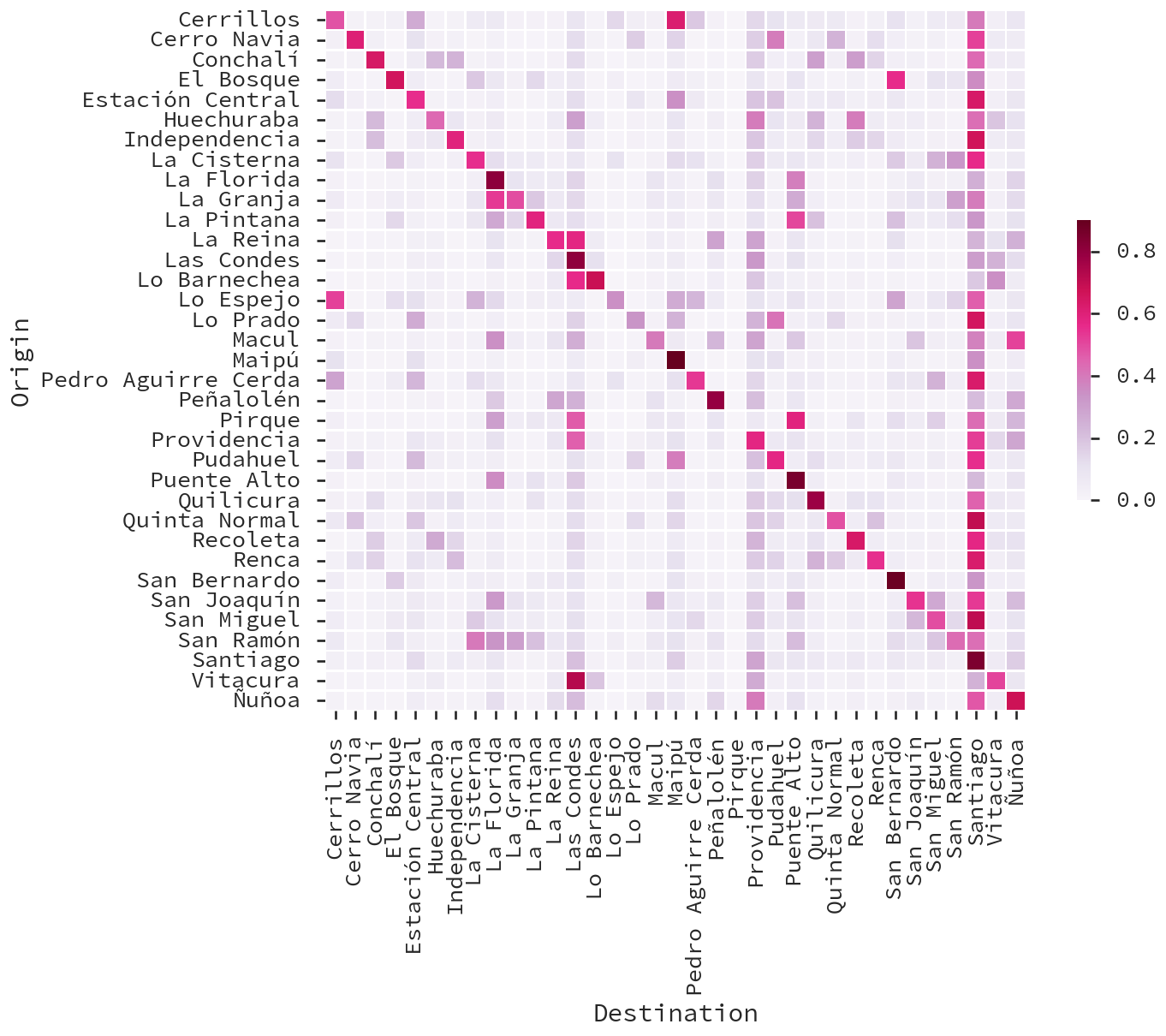}
\caption{Distributions of CDR trip data from Santiago, Chile. The matrix has been L2 normalized on rows (origins).}
\label{fig:cdr_matrix}
\end{figure}

After estimating the daily journeys from the graphical timetables, we
discarded users without trip and non-trip activities. Table
\ref{table:cdr_means} shows the final number of users considered (after
filtering by entropy, and after filtering without valid
trips/non-trips). From these activities, we built a transient OD matrix
for each day, using the initial and final positions of each trip, which
were assigned to their corresponding municipalities. Since we estimated
matrices for many days, we averaged the number of trips for each OD pair
of origin/destiny municipalities $(m_o, m_d)$. Figure
\ref{fig:cdr_matrix} shows the resulting matrix.

\begin{figure}[t]
\centering
\includegraphics[width=\linewidth]{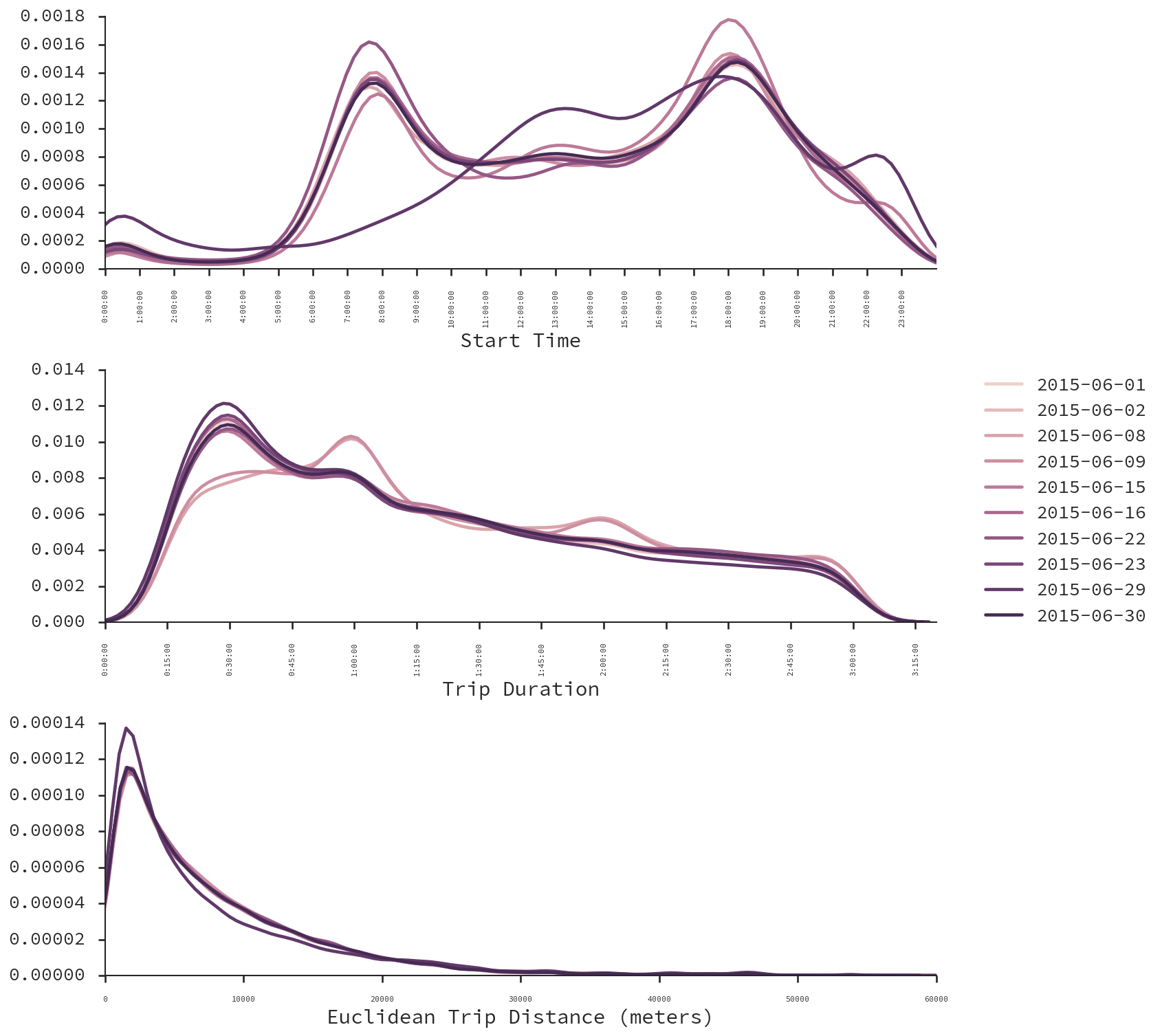}
\caption{Distributions of CDR trip data for each day.}
\label{fig:cdr_distributions}
\end{figure}

We also estimated the trip variables analyzed before for the ODS. Figure
\ref{fig:cdr_distributions} shows the kernel density estimations of each
distribution for each day. One can see that, overall, the distributions
are mostly similar for all days, with the following exceptions: start
time distribution is different on June 29th, and trip duration
distribution is different on June 8th and 9th. On June 9th there was a
strike in the Santiago public transport system, which explains partly
the increased trip time in comparison to the other days. Additionally,
on June 29th the semi-final of the latin-american soccer championship
\emph{Copa América}, where Chile was a contender, was played at 8pm. One
can see that the number of trips in the afternoon was much higher than
in the morning, making the morning peak to shrink in the density
estimation. Moreover, the distribution highlights that the afternoon
peak was earlier than usual, and that there was a night peak after the
soccer match.

Table \ref{table:cdr_means} shows the number of users (and their trips),
the mean trip duration, and the mean euclidean distance of trips for
each day. Mean times vary within 74.78 and 83.86 minutes, and mean
distances vary within 7.5 and 7.8 kilometers. One can see that both
distance and time are over-estimated in comparison to the ODS (mean time
41 minutes; and mean distance 6.05 kilometers). The differences in time
could appear due to the latency in antenna changes and to CDR event
granularity (which happens every 15 minutes in the case of data
connections). The differences in distance could be explained by the
approximation of each individual position to the corresponding antennas.
However, even though the means are different, the distributions have
similar shapes to those from the ODS, as displayed on Figure
\ref{fig:cdr_distributions}.

\spara{Comparisons with ODS} A key question is how much different our
results are with respect to the ODS. First, we estimated the Spearman
rank-correlation between our results and the ODS at municipal level,
obtaining $\rho = 0.89$ ($p < 0.001$). The correlation is very high,
which means that our averaged matrix reflects the flow of people in the
city very well. To compare to what extent our result is good, we refer
to a 2013 OD matrix estimated from the public transport smart-card data
\cite{munizaga2012estimation}.\footnote{\url{http://www.dtpm.cl/index.php/2013-04-29-20-33-57/matrices-de-viaje}}
This matrix has a correlation of $\rho = 0.3$ ($p < 0.001$) with the
ODS, a result that indicates that the ODS captures a diversity of trip
modes, not only public transport. We also tested what happened if we did
not consider the matrix diagonal (\ie, without intra-municipal trips),
and we observed that we mantain our correlation, while the public
transport OD increased ($\rho = 0.32$). This makes sense--short trips
are less likely to use public transport (\eg, if the destiny is at
walking distance).

\begin{figure}[t]
\centering
\includegraphics[width=\linewidth]{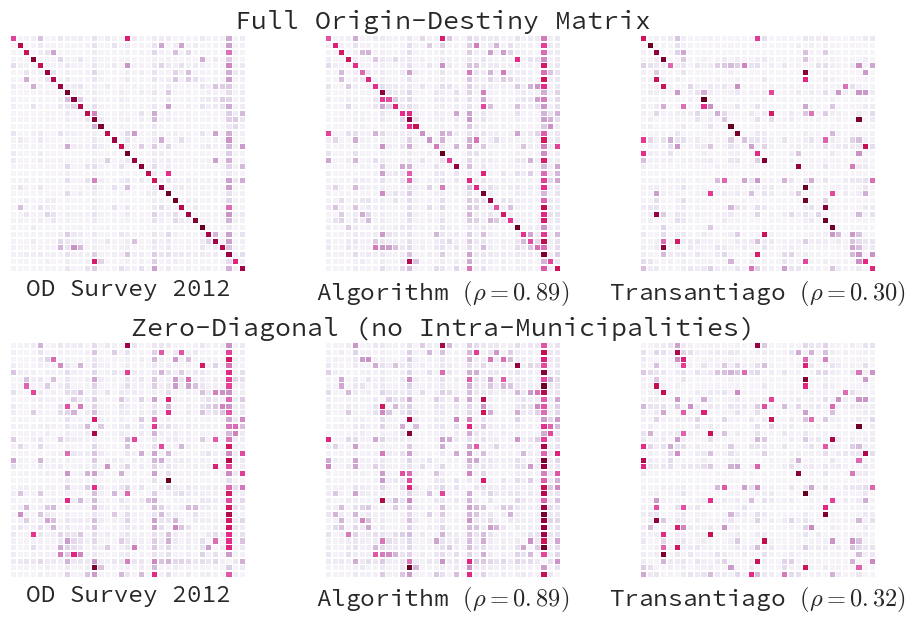}
\caption{Comparison of OD matrices: OD Survey, our method, and public transport \cite{munizaga2012estimation}.}
\label{fig:matrix_comparison}
\end{figure}

\begin{figure}[t]
\centering
\includegraphics[width=\linewidth]{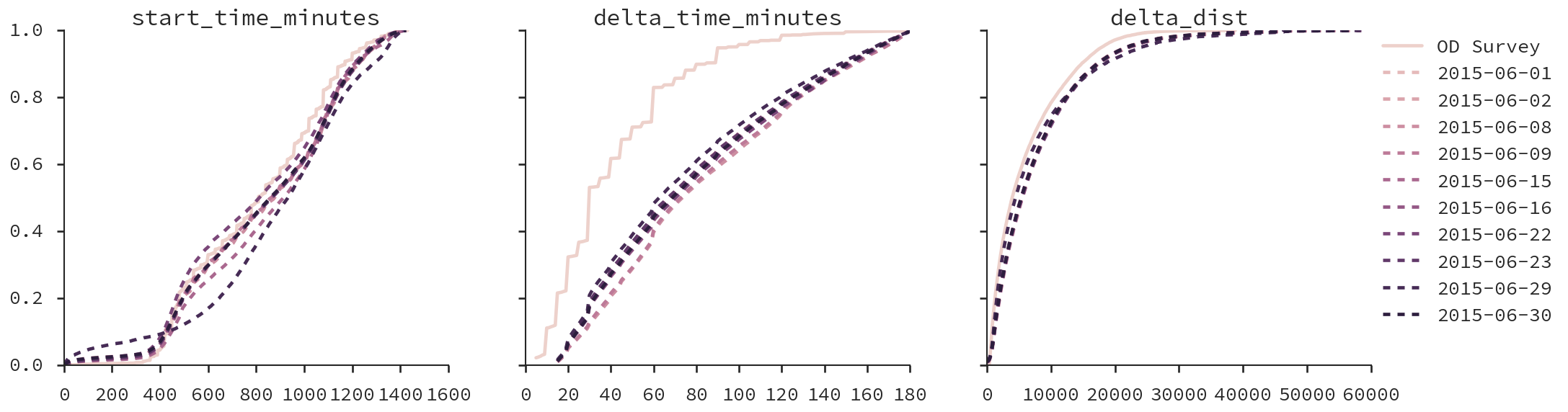}
\caption{Comparison between OD Survey and CDR-based data.}
\label{fig:cdr_eod_comparison}
\end{figure}

Finally, we visually compare whether the distributions of start time,
trip duration and trip distance are similar to those of the ODS. To do,
Figure \ref{fig:cdr_eod_comparison} shows the \emph{Cumulative Density
Functions} of these variables. We observe that, in terms of start time
and euclidean distance, the CDFs are very similar, although in terms of
trip duration there is a noticeable difference, as mentioned earlier.

\section{Discussion and Conclusions}\label{discussion-and-conclusions}

In this paper we predicted OD matrices from daily journeys built using
mobile data. We did so by proposing a geometric approach based on
graphical timelines, and found that, using a sample from mobile data
connectivity, we were able to reconstruct the OD pairs in a city, at
municipal level, with a very high correlation. Moreover, we were able to
estimate start time, trip time and trip distance, in a way that
resembled the ODS results, with exception of trip time, which needs
calibration to account for the delay in antenna change with respect to
the moment in which each trip started.

\spara{Implications} On the one hand, our algorithm is very simple and
can work with streaming data if what matters is the number of trips.
Consider the day in which a soccer match was played, and peak hours
shifted. Reportedly, the transport authorities did not account for this
shift, and instead they only created ad-hoc routes in public
transport.\footnote{\url{http://www.mtt.gob.cl/copaamerica/santiago}}.
Thus, our results could support urban computing applications
\cite{zheng2014urban} which need almost real-time transport data. On the
other hand, our method complements the ODS in important ways. The ODS is
performed every 10 years, but, as with any static picture, it does not
capture rich context-dependant dynamics of the city. Conversely, our
approach is more dense, as it can be applied even at daily scale to
observe differences that the ODS does not. In this paper we have shown
that, even working with a sample of anonymous data from mobile data, it
is possible to reconstruct part of the ODS, as well as finding diverging
days from the typical patterns, at a fraction of its costs. This can be
used to measure, for instance, what are the effects in urban flows
caused by transport measures like road space rationing.

\spara{Limitations and Future Work} Our geometric algorithm, with
arguably reasonable transport-based constraints, does not consider
information that could be useful in determining the turning points of
the day (\eg, land-use or census information). Additionally, the
distance and time estimation need to be corrected using scaling factors.
Then, in addition to address our limitations, the main line of research
for future work will be the characterization of non-trip activities. One
way of performing such characterization is through the analysis of
land-use derived from mobile data analysis
\cite{Graells-Garrido2015,nishi2014extracting}.

\spara{Acknowledgements} We thank Oscar Peredo, José García and Pablo
García for valuable discussion.